\documentclass[twocolumn,aps,amsmath,amyssymb,floatfix,pra]{revtex4}

\usepackage{graphicx}
\usepackage{dcolumn}
\usepackage{bm}
\begin{document}
\input{epsf.tex}

\title{Dispersion control for matter waves and gap solitons in optical superlattices}

\author{Pearl J.Y. Louis}
\author{Elena A. Ostrovskaya}
\author{Yuri S. Kivshar}

\affiliation{Nonlinear Physics Centre and ARC Centre of Excellence for
Quantum-Atom Optics, Research School of Physical Sciences and
Engineering, The Australian National University, Canberra ACT
0200, Australia}

\begin{abstract}
We present a numerical study of dispersion manipulation and formation of matter-wave gap solitons in a
Bose-Einstein condensate trapped in an optical superlattice. We
demonstrate a method for controlled generation of
matter-wave gap solitons in a stationary lattice by using an interference pattern of two condensate wavepackets, which  mimics the structure of the gap soliton near the edge of a
spectral band. The efficiency of this method is compared with that of
gap soliton generation in a moving lattice recently demonstrated
experimentally by Eiermann {\em et al.} [Phys. Rev. Lett., {\bf 92}, 230401 (2004)].  We show that, by changing the relative depths of
the superlattice wells, one can fine-tune the effective dispersion
of the matter waves at the edges of the mini-gaps of the
superlattice Bloch-wave spectrum and therefore effectively control both the
peak density and the spatial width of the emerging gap solitons.
\end{abstract}

\maketitle

\section{Introduction}

In recent years Bose-Einstein condensates (BECs) loaded into
optical lattices have become an important tool in the studies of
linear and nonlinear behavior of coherent waves in periodic
systems. The band-gap structure of the matter-wave spectrum
resulting from the periodicity of the optical trapping potential
is responsible for much of the observed coherent behavior of the
condensate, including Bloch oscillations and Landau-Zener
tunnelling~\cite{CristianiMorsch,Jona-LasinioMorsch}. One of the
more dramatic effects of the lattice on the BEC dynamics is the
dependence of the group-velocity dispersion of matter waves on the
curvature of the spectral band. The latter, in turn, is a function
of the wavepacket quasi-momentum in the lattice rest frame.  Due
to the possibility to create a moving optical lattice by varying the
relative detuning of the interfering laser beams, the condensate
can be loaded into the lattice with a well-defined quasi-momentum
\cite{DenschlagSimsarian}. This enables precise engineering of the
dispersion of the BEC wavepackets in moving optical lattices
~\cite{EiermannTreutlein,FallaniCataliotti}.

Dispersion management of a BEC wavepacket loaded into a
one-dimensional (1D) moving  lattice was recently investigated
experimentally~\cite{EiermannTreutlein,FallaniCataliotti,AnkerAlbiez},
in both linear (non-interacting) and non-linear (weakly
interacting)  regimes. In the nonlinear regime, a balance between
the effects of nonlinearity and dispersion can produce {\em a
bright matter-wave soliton}, a localized BEC wavepacket that
maintains a constant spatial structure. The creation of a bright
soliton in a \textit{repulsive} condensate (as opposed to an
attractive condensate) is \emph{only} possible due to the  negative effective dispersion at the edge of the first Brillouin zone (BZ) of a 1D optical lattice (see
e.g.~\cite{Pu}).  Driving the condensate from the middle to the edge of the
Brillouin zone achieves transition between the regimes of positive and negative effective dispersion. Evolution of the matter wavepacket in the negative dispersion regime can result in the
self-focusing of the repulsive BEC in the form of a fundamental
{\em gap soliton}~\cite{EiermannAnker}.

The absolute value of the dispersion
experienced by a BEC wavepacket can be varied by changing
the lattice depth~\cite{Pu,EiermannAnker}.
All previous {\em experimental} studies on matter-wave dispersion management and gap solitons  involved single-periodic {\em shallow} lattices
~\cite{EiermannTreutlein,FallaniCataliotti,AnkerAlbiez,EiermannAnker}
which are characterized by narrow spectral gaps, greater
curvatures of the spectral bands, and hence larger values of the
effective dispersion at the gap edges. Stronger dispersion
requires larger matter-wave nonlinearity, and hence larger atomic
density, and/or wider wavepackets to achieve the localized state
\cite{EiermannAnker}. Therefore, dispersion manipulation in an optical lattice can potentially deliver control over the characteristics of BEC solitons.

In this paper, we explore
novel possibilities to control the {\em magnitude} of dispersion
experienced by a BEC wavepacket {\em at the edges of the spectral
bands} by  modifying the shape of a double-periodic optical
\emph{superlattice}. As known from the theory of
diatomic lattices and semiconductor superlattices, an extra periodicity opens up additional narrow
stop-gaps (or mini-gaps) in the band-gap spectrum. Here we show
that the values of effective dispersion at the edges of these
mini-gaps can be varied within a much greater range than it is
possible for a single-period lattice of a reasonable depth.

To demonstrate the efficient dispersion control and generation
of {\em immobile} gap solitons, we investigate the process of
nonlinear localization in both moving and {\em stationary}
superlattices. In the former case, in order to access the gap
regions, the initially stationary condensate in the middle of the
ground-state Brillouin zone is driven to the band edge by
accelerating the lattice.  By the end of the {\em adiabatic}
acceleration process, the condensate wavepacket has a
quasi-momentum and an internal structure of a linear Bloch wave at
the gap edge.  In this paper, we numerically simulate this
process, starting from the initial adiabatic loading of the
condensate into a quasi-1D superlattice. Our numerical simulations
closely follow the current experimental procedure for matter-wave
gap-soliton generation~\cite{EiermannAnker}.

In addition, we suggest a new method for creating the correct
initial conditions for the soliton formation at the gap edge in a
{\em stationary} lattice, by interfering two identical wavepackets
with equal momenta of the opposite sign, corresponding to the opposite edges of the first
BZ.  This enables us to efficiently create a matter wavepacket with the
correct internal structure at the relevant gap edge and avoid large time scales associated with the adiabatic acceleration.  Such a wavepacket preparation technique was previously explored  in
optics~\cite{Feng,Skryabin,SukhorukovKivshar,Mandelik,Drago}. Here we show that this
method leads to more efficient soliton generation and shape
control of the emerging gap solitons via dispersion control in the
superlattice.

\section{Matter-wave spectrum in a superlattice}

In the mean-field approximation, the dynamics of a Bose-Einstein
condensate is described by the Gross-Pitaevskii (GP) equation. Assuming that the condensate cloud is elongated in the direction $x$, with the ratio of the corresponding frequencies at most $\Omega=\omega_{x}/\omega_{\perp} \sim 0.1$ \cite{EiermannAnker,FallaniCataliotti,Burger}, the condensate wavefunction in the axial dimension can be described by  the one-dimensional (1D) GP equation (see, e.g., Ref.~\cite{Perez-Garcia}):
\begin{equation}
\label{eq1DGPE} i\frac{\partial\psi}{\partial t} =
-\frac{1}{2}\frac{\partial^{2}\psi}{\partial x^{2}} + V(x,t)\psi +
g_{1D}|\psi|^{2}\psi,
\end{equation}
where  $V(x,t)$ is the external trapping potential, and  $g_{1D}$
characterizes the strength of the two-body interactions rescaled
for the case of the 1D geometry. The structure of the condensate
wavefunction in the transverse dimensions is determined by a tight
harmonic potential.

Assuming that any additional trapping along the axial direction
(e.g., due to a magnetic trap) is weak, we write $V(x,t)$ as a periodic potential of a 1D optical
superlattice:
\begin{equation}
\label{pot1DSL}
V(x)=U[\varepsilon\sin^{2}(K_{1}x)+(1-\varepsilon)\sin^{2}(K_{2}x)].
\end{equation}
The superlattice potential given by Eq.~(\ref{pot1DSL}) can be
obtained by superimposing two independent (either detuned from each other \cite{PeilPorto} or orthogonally polarized \cite{GrynbergRobilliard})  far off-resonance
single-periodic standing waves with different periods $d_{1}$ and
$d_{2}$. The larger of the two periods, e.g. $d=d_1$, defines the coarse periodicity of the lattice. The lattice wavevectors are given by $K_{1}=\pi/d_{1}$ and $K_{2}=\pi/d_{2}$ where the
commensurable periods are chosen such that $d_{1}/d_{2}=2$. The relative and total intensity of the standing waves are controlled by the parameters 
$0\leq\varepsilon\leq 1$ and $U$ respectively. 

Equation (\ref{eq1DGPE}), with the lattice potential given by (\ref{pot1DSL}), was made dimensionless by using the
characteristic length $a_{L}=d/\pi$, energy
$E_L=\hbar^{2}/ma_{L}^{2}=2E_{\rm rec}$, and time
$\omega^{-1}_{L}=\hbar/E_L$, where  $m$ is the atomic mass. In these
dimensionless units, and with the original three-dimensional condensate wavefunction normalized by $a^{-3/2}_L$, the nonlinear coefficient in Eq. (1) becomes:  $g_{1D}=2(a_s/a_L)(\omega_\perp/\omega_L)$. Here we use the parameters of the $^{87}$Rb
condensate: $m=1.44 \times 10^{-25}$~kg and $a_{s}=5.7$~nm. 
We assume that the periods of the standing waves forming the superlattice are
$d=d_{1}=350~\mathrm{nm}$ and $d_{2}=175~\mathrm{nm}$, so in the
dimensionless units, $K_1=1$ and $K_2=2$. With strong transverse
confinement of $\omega_\perp\sim 500-550
\mathrm{Hz}$~\cite{EiermannAnker,Burger,FortCataliotti},
$g_{1D}\sim 0.001$, and this is the value we use for most of the
simulations in this paper.

\begin{figure}
\setlength{\epsfxsize}{8.5 cm}
\begin{center}
\epsfbox{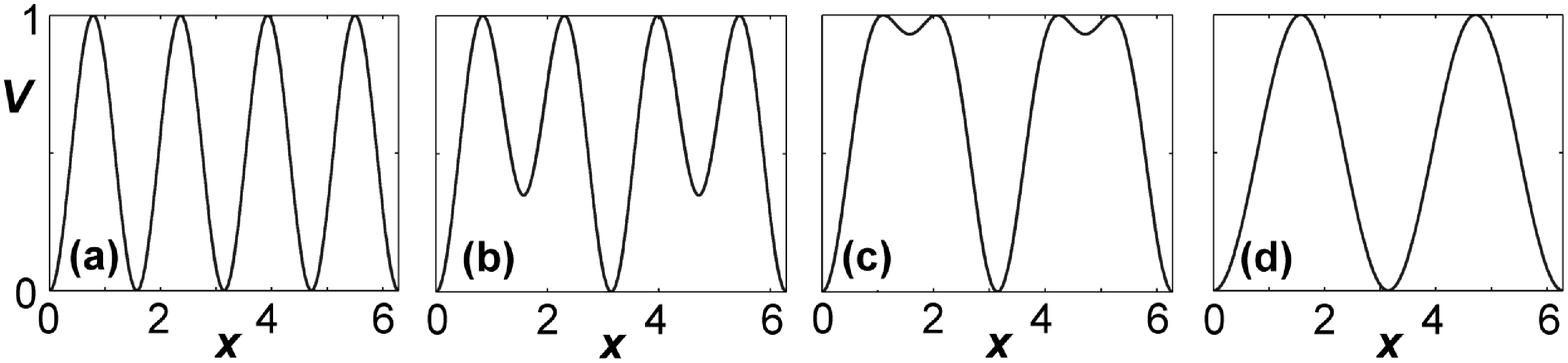}
\end{center}
\caption{\label{figmakingSL} The structure of the superlattice
potential described by Eq.~(\ref{pot1DSL}) for different values of
$\varepsilon$ and a fixed potential height $V_0=1$: 
single-periodic optical lattice for (a) $\varepsilon=0$ and (d) $\varepsilon=1$, and
double-periodic superlattices for (b) $\varepsilon=0.3$ and (c)
$\varepsilon=0.7$.}
\end{figure}

The shape of the optical superlattice
(\ref{pot1DSL}) depends on the values of the
parameters, $U$ and $0\leq\varepsilon\leq 1$. In the limits $\varepsilon \to
0$ or $\varepsilon \to 1$, the lattice becomes single-periodic,
and $U$ coincides with the height of the lattice $V_{0}$. For
$\varepsilon \neq 0$ and $\varepsilon \neq 1$, Eq.~(\ref{pot1DSL})
describes a double-periodic superlattice with $U$ defined through
the amplitude of the periodic potential $V_0$ as
\[
U=16 V_0 \frac{(1-\varepsilon)}{(4-3\varepsilon)^{2}}. \]
Figure~\ref{figmakingSL} shows several examples of the
superlattice potential (\ref{pot1DSL}) with the constant amplitude $V_0=1$.
As seen in the figure, the relative depth of the large and small lattice
wells can be manipulated by varying
$\varepsilon$, whilst keeping the height of the
lattice constant.  

\begin{figure}
\setlength{\epsfxsize}{8.7 cm}
\begin{center}
\epsfbox{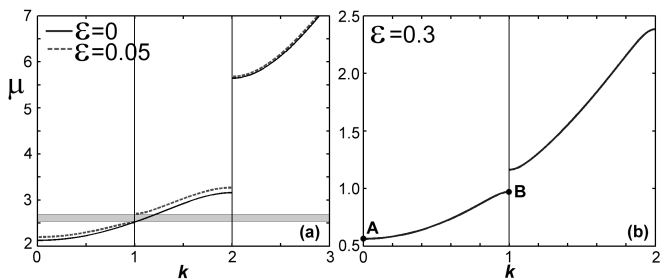}
\end{center}
\caption{\label{figbandgap} (a) The band-gap spectra for a
single-periodic optical lattice ($\varepsilon=0$) and a
double-periodic superlattice with weak modulation
($\varepsilon=0.05$).  In both cases, the lattice height is set to
$V_{0}=5$.  The first mini-gap is shaded. (b) The spectrum around
the first mini-gap for a strongly modulated superlattice at
$V_{0}=1$ and $\varepsilon=0.3$.}
\end{figure}

In the linear regime ($g_{1D}=0$), the stationary solutions to
Eq.~(\ref{eq1DGPE}) with the periodic potential are Bloch
waves:
\begin{equation}\label{BW}
\psi(x,t) =  \phi(x)e^{-i\mu t} = u(x) e^{ik(\mu) x} e^{-i\mu t},
\end{equation}
where $\mu$ is the matter-wave chemical potential and the function $u(x)$
has the periodicity of the lattice.
The linear matter-wave spectrum $\mu(k)$
consists of bands where the real wavenumbers $k(\mu)$ correspond
to the oscillatory Bloch-wave solutions. The bands are separated by ``gaps''
where ${\rm Im}(k)\neq 0$.  Figure~\ref{figbandgap}(a) shows the
band-gap diagrams plotted in the extended zone scheme for a
single-periodic lattice ($\varepsilon=0$) and a superlattice
($\varepsilon=0.05$), both for the lattice amplitude of $V_{0}=5$.
Since the coarse periodicity of the superlattice is twice that
of the single-periodic lattice, the size of the Brillouin zones of
the superlattice are half that of the single-periodic lattice at
$\varepsilon=0$. This leads to mini-gaps appearing in the
superlattice band-gap spectra at $k=1$.  A shaded stripe in Fig.~\ref{figbandgap}(a)
shows the lowest energy mini-gap at $\varepsilon=0.05$. As $\varepsilon$ grows, the size of the
mini-gap increases.

\begin{figure}
\setlength{\epsfxsize}{8.7 cm}
\begin{center}
\epsfbox{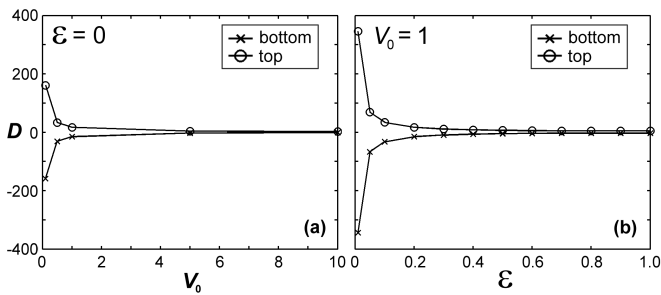}
\end{center}
\caption{\label{figDvsep}  The effective group-velocity dispersion
$D$ (a) at the edges of the first gap of a single-periodic optical
lattice of varying strength, and (b) at the edges of the first
mini-gap for a superlattice of height $V_{0}=1$ and period
$d=\pi$,  as the shape of the superlattice is changed by varying
$\varepsilon$.}
\end{figure}

For an interacting condensate ($g_{1D}\neq 0$), the solutions of
the model equation near the band (or mini-band) edges can be
sought in the form: $\psi(x,t)=\exp{(-i\mu t)}f(x,t)\phi(x)$,
where $\phi(x)$ is the linear Bloch wave at the corresponding band
edge and $f(x,t)$ is a slowly varying envelope~\cite{Sipe,SteelZhang,Pu}.
Then, the dynamics of the envelope are governed by the reduced GP equation~\cite{Sipe,SteelZhang,Pu}:
\begin{equation}
\label{eq1DGPEeffmass} i\frac{\partial f(x,t)}{\partial t} =
\left\{-\frac{D}{2}\frac{\partial^{2}}{\partial x^{2}}  +
\tilde{g}_{1D}|f(x,t)|^{2}\right\}f(x,t),
\end{equation}
where $D=\partial^{2}\mu/\partial k^{2}=\partial v_{g}/\partial k$
is the effective group velocity ($v_g$) dispersion and
$\tilde{g}_{1D}=g_{1D}\int|\phi(x)|^4dx/\int|\phi(x)|^2dx$. 

Figure~\ref{figDvsep}(a) shows the effective group velocity
dispersion, $D$, near the lowest energy gap of a single-period lattice for varying lattice height. As seen in
Fig.~\ref{figDvsep}(a), the dispersion is negative at the bottom
of the gap and positive at the top.  As the condition
$g_{1D}D<0$ is required for the formation of bright gap solitons,
bright gap soliton families originate near the bottom
of the gaps, for repulsive condensates ($g_{1D}>0$),
and near the top of the gaps, for attractive
condensates ($g_{1D}<0$)~\cite{Zobay,Alfimov,Pearl,Efremidis,Dima}, and
exist for the entire range of chemical potentials within that gap \cite{Alfimov,Pearl,Efremidis}.
The group velocity $v_{g}=
\partial\mu/\partial k$ vanishes at the band edges, hence the
gap solitons form as immobile localized wavepackets. We note that formation and dynamics of gap solitons could be significantly modified in the case when the harmonic confinement in the longitudinal dimension becomes important \cite{morsch,adhikari,chaos}. However, provided the longitudinal harmonic trapping is weak, the band structure imposed by the lattice is still well defined. Moreover, in many experiments the longitudinal trapping potential is either sufficiently flat on the scale of hundreds of lattice wells, or is removed altogether after the condensate is loaded into an optical lattice \cite{EiermannAnker}.

Gap soliton formation can also occur in the minigaps of the superlattice spectrum.  The size of the mini-gap increases with growing $\varepsilon$. This decreases the curvature in the band structure immediately surrounding the mini-gap, leading to a decrease in the magnitude
of the effective dispersion at the gap edges.
Figure~\ref{figDvsep}(b) shows that simple variation of the relative depths of the two wells
in the superlattice (i.e. $\varepsilon$) at a constant lattice
height ($V_{0}=1$)  opens up access to a large range of effective dispersions at the edges of the
mini-gap. As Figure~\ref{figDvsep}(a) shows, this kind of dispersion control can also be
performed in a single-periodic lattice by changing the lattice
height. However, the range
of effective dispersion values in a single-periodic
lattice is much smaller than that achievable in a superlattice,
and the larger proportion of this range is only accessible for
$V_0< 1$.

\begin{figure}
\setlength{\epsfxsize}{8 cm}
\begin{center}
\epsfbox{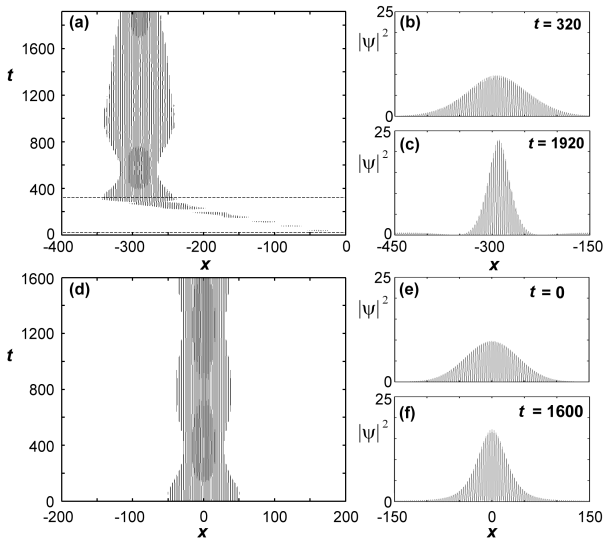}
\end{center}
\caption{\label{figsinglewp} Soliton generation by (a) a single
Gaussian wavepacket accelerated to the edge of the first mini-gap
($k=1$) in a moving superlattice and (d) interference of two counter-propagating Gaussian
wavepackets in a stationary superlattice.  Both superlattices have $\varepsilon=0.3$ and
$V_{0}=1$.  Figure (a) is plotted in the lattice frame moving at
$v_{\rm rec}$.  Initially the wavepacket contains $500$ atoms, and
has a width parameter of $w=100$.  In (d) the initial parameters
are $w=81.6$ and $N=500$. (b,c) and (e,f) capture the spatial
structure of the wavepackets at different evolution times.}
\end{figure}

\section{Gap-soliton generation in a moving lattice}

The key to producing a spatial gap soliton in a BEC is
to access the band-gap edge of the linear spectrum, where the
suitable effective dispersion can assist spatial self-focusing of the
coherent matter waves.  For a repulsive condensate, this occurs at the
bottom edge of the first gap, where the dispersion is negative. In
experiments, the stationary condensate is initially loaded into
the lattice ground state by slowly ramping up the lattice
height. This produces a wavepacket in the lowest energy band
($n=0$) in the middle of the first Brillouin zone (centered at
$k=0$).  To ensure that other energy bands are not populated, the
ramping process should be adiabatic.  To adiabatically load the
condensate into the lattice at quasi-momentum $k$, the change of
the single-particle BEC Hamiltonian on the time scale of a single
Rabi oscillation associated with the transition between the Bloch
states in the ground state, $\phi_{0,k}$, and the first excited
band, $\phi_{1,k}$, should be much smaller than the energy gap
between the $n=0$ and $n=1$ bands at quasi-momentum $k$. In our
dimensionless units, this adiabaticity condition translates to
$\Delta V_0/t_{\rm R}\ll \Delta \mu^2(k)$
(\cite{DenschlagSimsarian,PeikBenDahn,DudarevDiener}), where
$t_{\rm R}$ is the duration of the lattice ramp-up, $\Delta
V_0$ is the total change of the lattice amplitude, and $\Delta
\mu(k)$ is the size of the first (mini)gap at the quasi-momentum
$k$. In the middle of the Brillouin zone (BZ), at $k=0$, the size of the first
mini-gap is equal to four single-photon recoil
energies~\cite{DenschlagSimsarian}. Therefore, the single-particle
adiabaticity condition for loading a stationary condensate into
the lattice at $k=0$ is, $\Delta V_0/t_{\rm R}\ll \Delta \mu^2(k=0)=4$.

In order to access  the band edge, a constant acceleration is
introduced {\em suddenly}~\cite{DenschlagSimsarian}, by applying
time-dependent frequency detuning to the lasers used to create the
lattice. The superlattice potential given by Eq.(\ref{pot1DSL})
becomes time-dependent: $V(x,t)=U[\varepsilon\sin^{2}(K_1(x -
\delta t))+(1-\varepsilon)\sin^{2}(K_2(x - \delta t))]$. Then, the reference
frame in which the lattice is stationary (the lattice rest frame) is
moving in the laboratory frame with velocity $v(t)=\delta(t)$.  If
the acceleration, $a=d\delta/dt$, is constant, a constant and uniform force
 is applied to the condensate~\cite{PeikBenDahn}, which is
somewhat analogous to the application of a constant homogeneous electric
field to electrons in a crystal. 
The quasi-momentum of the condensate depends on the duration of the
acceleration, $t_{\rm A}$, as $k(t_A) = k(0) + at_{\rm A}$, and 
in order to move the condensate to the band-edge ($k(t_{\rm
A})=1$), the recoil velocity
$v_{\rm rec}=1$ of the lattice is reached with acceleration $a=1/t_{\rm A}$. The acceleration is
adiabatic, i.e. the upper bands are not populated, provided the
probability of transitions between the $n=0$ and $n=1$ bands are
low.  This requires a small magnitude of the acceleration, $|\Delta \delta/t_{\rm A}| \ll \Delta \mu(k)$
~\cite{PeikBenDahn}.  For acceleration at a constant lattice height, the critical energy gap is at the edge of the Brillouin zone ($k=\pm 1$), where the gap is generally much smaller than that at $k=0$, as can
be seen in Fig.~\ref{figbandgap}.   The ramping adiabaticity condition is also much harder to fulfil at the
band-gap edge than in the middle of the Brillouin zone. This is why it is
preferable to load the condensate into the lattice at $k=0$ and
then accelerate it to the band-edge instead of simply ramping up
an already moving lattice at velocity $v_{\rm rec}$.

We model the process of ramping and accelerating the lattice by
starting with an initially stationary Gaussian wavepacket:
$\psi(x,t=0)=A\mathrm{exp}(-x^{2}/w^{2})$ where $A$ is the
amplitude and $w$ is the width. 
Figure~\ref{figsinglewp}(a) shows the three stages of the
evolution of an initially Gaussian wavepacket with $w=100$ and
$N=500$ atoms in a superlattice with $\varepsilon=0.3$ and a
potential depth of $V_{0}=1$ [the band-gap structure of this
lattice is shown in Fig.~\ref{figbandgap}(b)]:

\begin{itemize}

\item

The superlattice is ramped up from $V_{0}=0$ to $V_{0}=1$ in time
$t_{\rm R}=20$.  This places the wavepacket at $k=0$ in the
lowest energy band [point A in Fig.~\ref{figbandgap}(b)]. The
wavepacket develops an internal structure with the same spatial
properties as a Bloch wave at $k=0$.

\item

The superlattice is accelerated to the edge of the first mini-gap
[point B in Fig.~\ref{figbandgap}(b)] in time $t_{\rm A}=300$.

\item

The wavepacket is then left to evolve in a superlattice moving
with constant velocity $v_{\rm rec}$ i.e. at the edge of the
Brillouin zone. Figures~\ref{figsinglewp}(a-c) are plotted in a
frame moving at the recoil velocity $v_{\rm rec}$ so that the
wavepacket appears stationary when it is at the gap edge where its
group velocity with respect to the lattice is zero.

\end{itemize}

For sufficiently long  times of the wavepacket evolution at the
mini-band edge, a localized wavepacket with the spatial structure
of a gap soliton emerges, as seen in Fig.~\ref{figsinglewp}(c). A
fundamental gap soliton near a gap edge takes the form of a broad
sech-shaped envelope with an internal structure resembling the
linear Bloch waves at that particular edge.  This spatial
structure allows sufficiently broad gap solitons to be described
using a wave-envelope approximation with an appropriately chosen
carrier wave, as described in the previous section. Due to large
differences between the ideal shape of the soliton and the shape
of the initial wavepacket, significant oscillations in the shape of the wavepacket are observed [see
Fig.~\ref{figsinglewp}(a)]. This is also a well known part of soliton
formation in the absence of the lattice~\cite{Agrawal,Burak}. The closer the initial width of the wavepacket is to
 that of the soliton, the more efficient is the process of
soliton generation. This is illustrated in Figs.~\ref{figsinglewp}(d-f), where a slightly narrower initial wavepacket was used. As a general guide, the peak density and width of the initial wavepacket required for the formation of the fundamental gap soliton are related as: $A^2 \sim |D|/(w^2\tilde{g}_{1D})$ (see \cite{Agrawal}, ch. 5). While the time scale in
Figs.~\ref{figsinglewp}(a,d)  is not sufficient to see the shape of
the wavepacket stabilize, as we evolve the wavepacket even further, it is clear that the oscillations are gradually being damped as excess atoms are removed into the
low-density background.

The spatial extent of the initial  wavepacket is required to be
large ($\sim 90$ lattice sites), so that the momentum distribution
is very narrow compared to the size of the Brillouin zone.  The
acceleration process works best with a narrow momentum
distribution and hence a wide wavepacket, which is broader than
the soliton that is eventually formed.  If the momentum
distribution is too broad, undesirable effects may occur, e.g. the
wavepacket can develop a strong asymmetry as it passes through the
zero dispersion point, as described in
Ref.~\cite{EiermannTreutlein}.   A wide initial wavepacket also
reduces the maximum density in the initial wavepacket used to form
a fundamental soliton since the width of the fundamental soliton's
envelope near the band edge is inversely proportional to its
amplitude, $A\sim 1/w$~\cite{Agrawal}.

For both the ramping and acceleration processes, we have checked
the momentum components of the wavepacket to ensure that there is
no significant excitation of the upper bands.  As discussed above,
the rate at which the lattice can be adiabatically accelerated
depends very strongly on the size of the energy gap. The size of the mini-gaps
for superlattices with $\varepsilon \ll 1$ are much smaller than
the size of the gaps of a single-periodic lattice for the same
lattice height.  This makes the above described techniques for
condensate preparation  at the edges of a superlattice mini-gap
rather difficult to use due to the large acceleration times. 

Difficulty in accelerating the condensate to the gap edge for
narrow gaps produced in shallow single-periodic
optical lattices was also noted in recent
experiments~\cite{EiermannAnker}. However, smaller gaps are useful
because they potentially enable greater values of the effective
dispersion [see Fig.~\ref{figDvsep}(a)].
Below, we study an alternative scheme for gap-soliton
generation in a {\em stationary} lattice. Since the problems of
large time-scales associated with conditions of adiabatic ramping
and acceleration are avoided in this case, the method would be
equally useful for low modulation height single-periodic optical
lattices and superlattices with narrow mini-gaps.

\section{Gap-soliton generation in a stationary lattice}

An alternative method for creating a wavepacket with the correct
quasi-momentum and internal structure to produce {\em spatial gap
solitons} was first suggested theoretically in the context of nonlinear
optics~\cite{Feng,Skryabin,SukhorukovKivshar}, and was recently
employed in experiments on weakly coupled waveguide arrays and
optically-induced photonic lattices~\cite{Mandelik,Drago}. Applied
to matter waves in optical lattices, this technique means that, instead of using a
moving lattice to gradually drag a BEC wavepacket  to
the edge of the Brillouin zone, we start with two non-stationary
wavepackets with opposite momenta  (i.e. $k=1$ and 
$k=-1$) corresponding to the Bragg
reflection condition in a {\em stationary} lattice. Their interference produces a  
matter wave with the internal structure resembling that of the Bloch wave at the gap edge.
This is expected since the Bloch wave at the edge of a gap is a
periodic standing wave formed at the Bragg reflection condition
and hence can be presented as the superposition of identical
forward and backward moving travelling waves.  Recently this
method was used to theoretically demonstrate the formation of optical spatial
gap solitons in binary waveguide arrays, the nonlinear optical
analogue to our particular optical superlattice
system~\cite{SukhorukovKivshar} for a focusing (attractive)
nonlinearity.  In Ref.~\cite{SukhorukovKivshar} it was found that
this method allows for the efficient production of spatial gap
solitons at the top of the mini-gap. Here we apply this method to
{\em repulsive} condensates (and hence the soliton formation at the \emph{bottom} of the
mini-gap).

\begin{figure}
\setlength{\epsfxsize}{8 cm}
\begin{center}
\epsfbox{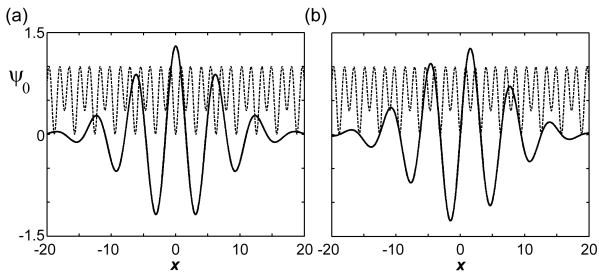}
\end{center}
\caption{\label{figInterference} Interference patterns of two
Gaussian wavepackets as described by Eq.~(\ref{eq2wp}) with
$A=1.3$, $k=1$, $w=10$, and (a) $\theta_{1}=0$, $\theta_{2}=0$,
and (b) $\theta_{1}=3\pi/2$, $\theta_{2}=\pi/2$. Dotted lines show
the superlattice potential  ($\varepsilon=0.3$ and $V_{0}=1$).}
\end{figure}

We assume the initial state of condensate evolution in the form of two
identical Gaussian wavepackets, $\psi_{1}(x,t=0)$ and
$\psi_{2}(x,t=0)$, with momenta $k_{1}=k$ and $k_{2}=-k$ and
phases $\theta_{1}$ and $\theta_{2}$ respectively.  These
wavepackets could be created using Bragg scattering
techniques. Assuming that both wavepackets are initially centered at the
origin, their superposition is given by the expression:
\begin{equation}
\label{eq2wp} \psi(x,t=0)=Ae^{-x^{2}/w^{2}}\left[ e^{ikx+i\theta_{1}}+e^{-ikx+i\theta_{2}} \right],
\end{equation}
where $A$ is a constant amplitude and $w$ is a width. By setting
the correct $k$, $\theta_{1}$ and $\theta_{2}$, the symmetry of
$\psi(x,t=0)$ can be matched to that of the Bloch waves at various
gap edges. For example, with $k=1$, $\theta_{1}=\theta_{2}=0$,
Eq.~(\ref{eq2wp}) simplifies to $\psi(x,t=0)=A\exp (-x^2/w^2)\cos
(x)$, which has the spatial structure of the Bloch wave at the
bottom edge of the first mini-gap of a superlattice potential
[see, e.g., Fig.~\ref{figInterference}(a)].  Setting $k=1$,
$\theta_{1}=3\pi/2$, and $\theta_{2}=\pi/2$ gives $\psi(x,t=0)=-A
\exp (-x^2/w^2)\sin(x)$ which mimics the Bloch wave at the top
edge of the first mini-gap [see, e.g.,
Fig.~\ref{figInterference}(b)].

\begin{figure}
\setlength{\epsfxsize}{8 cm}
\begin{center}
\epsfbox{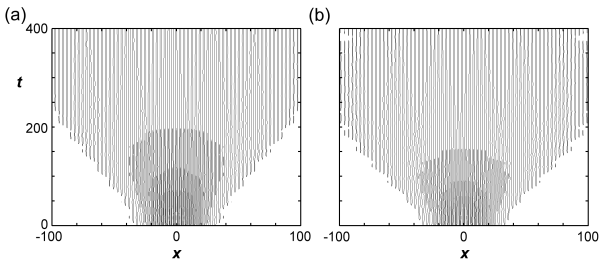}
\end{center}
\caption{\label{figlinear} Linear ($g_{1D}=0$) evolution of
wavepackets formed by the interference of two Gaussian wavepackets
described by Eq.~(\ref{eq2wp}) in a superlattice potential
($\varepsilon=0.3$ and $V_{0}=1$).  Number of atoms is $N=1000$,
$k=1$, and $w=30$.  In (a) the phases of the two wavepackets are
$\theta_{1}=0$ and $\theta_{2}=0$.  In (b) the phases are
$\theta_{1}=3\pi/2$ and $\theta_{2}=\pi/2$, respectively.}
\end{figure}

\begin{figure}
\setlength{\epsfxsize}{8.6 cm}
\begin{center}
\epsfbox{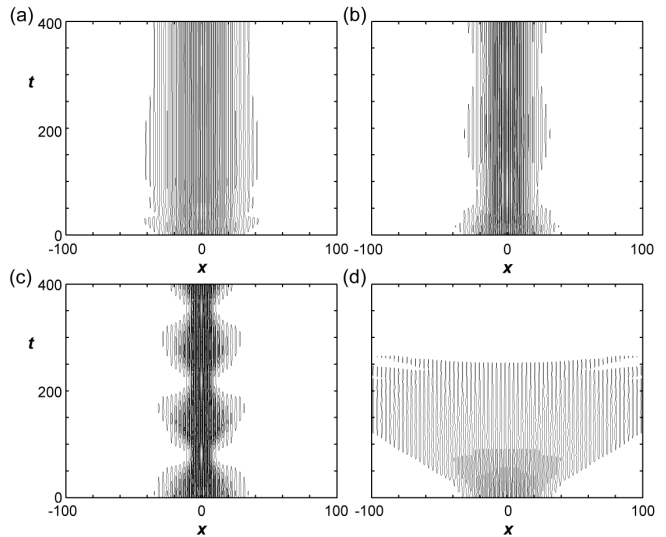}
\end{center}
\caption{\label{figRepulsive} Evolution of wavepackets described
by Eq.~(\ref{eq2wp})  with repulsive interaction $g_{1D}=0.001$,
$N=1000$, $k=1$, and $w=30$. In (a)-(c) $\theta_{1}=0$ and
$\theta_{2}=0$ i.e. the wavepacket is at the bottom of the first
mini-gap.  In (d) $\theta_{1}=3\pi/2$ and $\theta_{2}=\pi/2$ i.e.
the wavepacket is at the top of the first mini-gap.  Each of the
superlattice potentials have the same period and lattice height
$V_{0}=1$. The shape of each superlattice however varies: (a)
$\varepsilon=0.2$, (b) and (d) $\varepsilon=0.3$, (c)
$\varepsilon=0.5$.}
\end{figure}

In what follows we study the generation of matter-wave gap
solitons in the two cases discussed above.  That is, (a) $k=1$,
$\theta_{1}=0$ and $\theta_{2}=0$ i.e. at the bottom of the first
superlattice mini-gap and (b) $k=1$, $\theta_{1}=3\pi/2$ and
$\theta_{2}=\pi/2$ i.e. at the top of the first superlattice
mini-gap. We consider wavepackets containing  $N=1000$ atoms, with
the width of the spatial envelope set at $w=30$ which means the
initial wavepacket occupies around 30 wells.  With this soliton
excitation method, we can use wavepackets with small widths that
are closer to the fundamental soliton width at the corresponding
atom number, as a small initial momentum distribution is not
crucial.  If the initial wavepacket has a smaller width, then we
can use a larger nonlinearity, i.e. larger peak density, and
still obtain a fundamental soliton~\cite{Agrawal}. Using
Eq.~(\ref{eq2wp}) as the initial condition, we solve the 1D
evolution equation~(\ref{eq1DGPE})  with the stationary superlattice
potential~(\ref{pot1DSL}), by employing a Fourier split-step method implemented using
the code generator XMDS~\cite{xmds} 

In the absence of nonlinearity ($g_{1D}=0$), we expect that the
initial wavepacket disperses and spreads out.
Figure~\ref{figlinear} shows this occurring for both cases (a) and
(b) in a superlattice with $\varepsilon=0.3$ and $V_{0}=1$. The
wavepacket at the top of the mini-gap [Fig.~\ref{figlinear}(b)]
exhibits slightly greater dispersion then the wavepacket at the
bottom of the mini-gap [Fig.~\ref{figlinear}(a)], which agrees
with the values of the effective dispersion coefficient shown in
Fig.~\ref{figDvsep} .

\begin{figure}
\setlength{\epsfxsize}{8 cm}
\begin{center}
\epsfbox{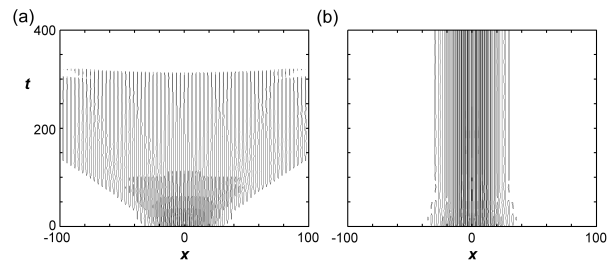}
\end{center}
\caption{\label{figAttractive} Evolution of a wavepacket formed by
the interference of two Gaussian wavepackets as described by
Eq.~(\ref{eq2wp}) in a superlattice potential with $V_{0}=1$ and
$\varepsilon=0.3$ for $N=1000$, $k=1$, and $w=30$. In contrast to
Figs.~\ref{figRepulsive}(b) and (d), we use an attractive nonlinearity,
$g_{1D}=-0.001$.  In (a) $\theta_{1}=0$ and $\theta_{2}=0$ i.e.
the wavepacket is at the bottom the first mini-gap.  In (b)
$\theta_{1}=3\pi/2$ and $\theta_{2}=\pi/2$ so that the wavepacket
is at the top of the first mini-gap.}
\end{figure}

When a repulsive nonlinearity ($g_{1D}>0$) is "turned on", the
interplay between negative dispersion and positive nonlinearity
($g_{1D}D<0$) results in the localization of the initial
wavepacket  into a gap soliton at the bottom of the mini-gap [see
Figs.~\ref{figRepulsive}(a-c)].  In contrast, at the top of the
mini-gap the repulsive nonlinearity interacts with the positive
dispersion ($g_{1D}D>0$) to accelerate the spreading and breakup
of the BEC wavepacket [see Fig.~\ref{figRepulsive}(d)]. When an
attractive nonlinearity ($g_{1D}<0$) is used instead, the
wavepacket forms a gap soliton at the top of the mini-gap while
the one at the bottom breaks up more quickly [see
Fig.~\ref{figAttractive}].  In the three cases shown in
Figs.~\ref{figRepulsive}(a-c), the fraction of atoms radiated from
the wavepacket during the process of gap-soliton formation
  is less than 10\%. In addition, the
localization occurs over time intervals that are almost an
order of magnitude shorter than those for the transient dynamics
of the wavepackets in the moving lattices.

\begin{figure}
\setlength{\epsfxsize}{8.5 cm}
\begin{center}
\epsfbox{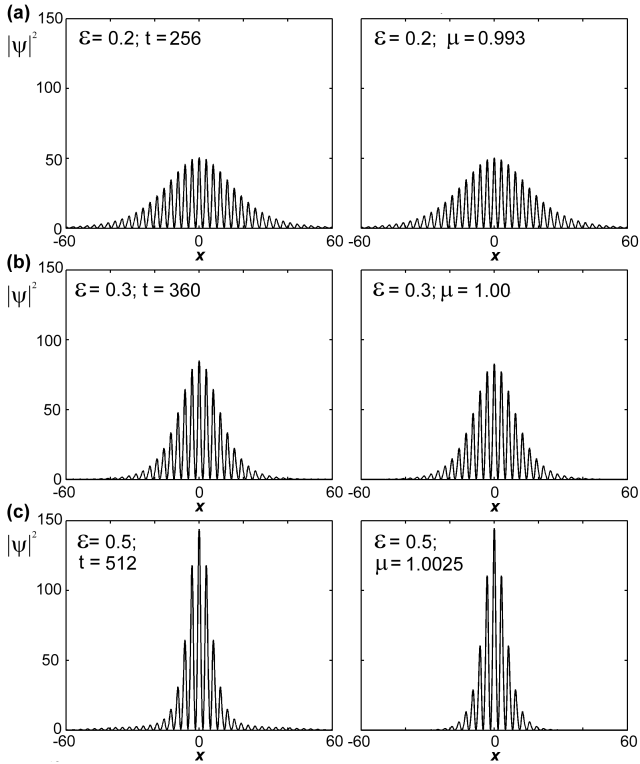}
\end{center}
\caption{\label{figCompare} Left: Density profiles of the pulses
shown in Figures~\ref{figRepulsive}(a)-(c). The initial conditions
are identical in each case, but the superlattice potential varies
with (a) $\varepsilon=0.2$, (b) $\varepsilon=0.3$ and (c)
$\varepsilon=0.5$.  In each case the lattice height is $V_{0}=1$.
Right: Exact stationary gap soliton solutions obtained by
numerically solving the time-independent version of the full model equation (1) for the same superlattice parameters as on the left.}
\end{figure}

In general, the interference method of gap-soliton generation
allows for more precise control of the initial wavepacket widths,
uninhibited by any requirements on the initial momentum
distribution. This allows us to perform an accurate analysis of the
effects of dispersion on the BEC localization process in the
superlattice potential by creating an initial wavepacket at the
band-gap edge with exactly the same parameters for each
superlattice. The effects of the dispersion control are
demonstrated in Figs.~\ref{figRepulsive}(a-c), which show the
gap-soliton generation in a mini-gap for three different
shapes of the superlattice potentials and the same repulsive nonlinearity ($g_{1D}=0.001$).  In each case, the lattice period and height are kept constant, as are the
parameters of the initial wavepacket.  The superlattice parameter
$\varepsilon$ varies and with it the relative well-depths of the
small and large wells in the superlattice potential.  With growing
$\varepsilon$, the generated gap soliton becomes narrower and the
average peak density increases.  The dramatic change in the shape of
the output soliton with variations in $\varepsilon$ is due to the
variations in the effective dispersion with $\varepsilon$, as seen in Fig.~\ref{figDvsep}(b).

Figure~\ref{figDvsep} shows that as $\varepsilon\rightarrow 1$ the
magnitude of the effective dispersion at the edges of the
superlattice band-gap decreases due to the increasing size of the
mini-gap.  Consequently, the peak
particle density required to form a soliton with
the {\em same number of atoms} needs to increase to compensate.
In Fig.~\ref{figCompare} we show the density profiles of the
pulses shown in Figs.~\ref{figRepulsive}(a-c) compared with the
{\em stationary} fundamental soliton solutions obtained numerically by
solving the time-independent GP equation for
identical superlattice parameters. Because of the transient
oscillations in the peak density of the dynamically generated gap
solitons, we use the average value over time for this comparison.
In each case, the matter-wave chemical potential is around the
value $\mu=1.0$, which is close to the bottom edge of the mini-gap
for each of the three values of $\varepsilon$.

\section{Conclusions}

In this paper, we have investigated an effective method of
generating matter-wave gap solitons - nonlinear localized states of a repulsive
Bose-Einstein condensate in a {\em stationary} optical lattice. We
demonstrated that, by interfering two matter waves, one can create a
wavepacket located at the gap edge with the appropriate internal
structure suitable for gap soliton formation.  The initial wavepacket prepared in this fashion can be better matched to the shape of the fundamental gap soliton, which leads to a more efficient and faster soliton
formation compared to the standard technique of lattice
acceleration. This method would be especially useful for small
energy gaps, e.g. in the case of superlattice mini-gaps and for
extremely shallow single-periodic optical lattices, where
preparation of the condensate wavepackets at a gap edge by using
lattice acceleration is experimentally difficult and time
consuming.  Using the interference method to simulate the gap soliton formation in
double-periodic optical superlattices, we have shown that it is
possible to fine-tune the effective dispersion of matter waves by
changing the shape of the superlattice rather than its height or
periodicity, and hence tailor the properties of the BEC
solitons.

\section{Acknowledgements}

This work has been supported by the Australian Research Council (ARC).
The authors thank
Andrey Sukhorukov, Markus Oberthaler, Dmitry Pelinovsky, and
Dmitry Skryabin for useful discussions and suggestions.


\begin{thebibliography}{29}
\expandafter\ifx\csname natexlab\endcsname\relax\def\natexlab#1{#1}\fi
\expandafter\ifx\csname bibnamefont\endcsname\relax
  \def\bibnamefont#1{#1}\fi
\expandafter\ifx\csname bibfnamefont\endcsname\relax
  \def\bibfnamefont#1{#1}\fi
\expandafter\ifx\csname citenamefont\endcsname\relax
  \def\citenamefont#1{#1}\fi
\expandafter\ifx\csname url\endcsname\relax
  \def\url#1{\texttt{#1}}\fi
\expandafter\ifx\csname urlprefix\endcsname\relax\def\urlprefix{URL }\fi
\providecommand{\bibinfo}[2]{#2}
\providecommand{\eprint}[2][]{\url{#2}}

\bibitem[{\citenamefont{Cristiani et~al.}(2002)\citenamefont{Cristiani, Morsch,
  M{\"{u}}ller, Ciampini, and Arimondo}}]{CristianiMorsch}
\bibinfo{author}{\bibfnamefont{M.}~\bibnamefont{Cristiani}},
  \bibinfo{author}{\bibfnamefont{O.}~\bibnamefont{Morsch}},
  \bibinfo{author}{\bibfnamefont{J.~H.} \bibnamefont{M\"uller}},
  \bibinfo{author}{\bibfnamefont{D.}~\bibnamefont{Ciampini}}, \bibnamefont{and}
  \bibinfo{author}{\bibfnamefont{E.}~\bibnamefont{Arimondo}},
  \bibinfo{journal}{Phys. Rev. A} \textbf{\bibinfo{volume}{65}},
  \bibinfo{pages}{063612} (\bibinfo{year}{2002}).

\bibitem[{\citenamefont{Jona-Lasinio et~al.}(2003)\citenamefont{Jona-Lasinio,
  Morsch, Cristiani, M\"uller, Courtade, Anderlini, and
  Arimondo}}]{Jona-LasinioMorsch}
\bibinfo{author}{\bibfnamefont{M.}~\bibnamefont{Jona-Lasinio}},
  \bibinfo{author}{\bibfnamefont{O.}~\bibnamefont{Morsch}},
  \bibinfo{author}{\bibfnamefont{M.}~\bibnamefont{Cristiani}},
  \bibinfo{author}{\bibfnamefont{N.}~\bibnamefont{Malossi}},
  \bibinfo{author}{\bibfnamefont{J.~H.} \bibnamefont{M\"uller}},
  \bibinfo{author}{\bibfnamefont{E.}~\bibnamefont{Courtade}},
  \bibinfo{author}{\bibfnamefont{M.}~\bibnamefont{Anderlini}},
  \bibnamefont{and} \bibinfo{author}{\bibfnamefont{E.}~\bibnamefont{Arimondo}},
  \bibinfo{journal}{Phys. Rev. Lett.} \textbf{\bibinfo{volume}{91}},
  \bibinfo{pages}{230406} (\bibinfo{year}{2003}).

\bibitem[{\citenamefont{Denschlag et~al.}(2002)\citenamefont{Denschlag,
  Simsarian, H{\"{a}}ffner, McKenzie, Browaeys, Cho, Helmerson, Rolston, and
  Phillips}}]{DenschlagSimsarian}
\bibinfo{author}{\bibfnamefont{J.~H.} \bibnamefont{Denschlag}},
  \bibinfo{author}{\bibfnamefont{J.~E.} \bibnamefont{Simsarian}},
  \bibinfo{author}{\bibfnamefont{H.}~\bibnamefont{H{\"{a}}ffner}},
  \bibinfo{author}{\bibfnamefont{C.}~\bibnamefont{McKenzie}},
  \bibinfo{author}{\bibfnamefont{A.}~\bibnamefont{Browaeys}},
  \bibinfo{author}{\bibfnamefont{D.}~\bibnamefont{Cho}},
  \bibinfo{author}{\bibfnamefont{K.}~\bibnamefont{Helmerson}},
  \bibinfo{author}{\bibfnamefont{S.~L.} \bibnamefont{Rolston}},
  \bibnamefont{and} \bibinfo{author}{\bibfnamefont{W.~D.}
  \bibnamefont{Phillips}}, \bibinfo{journal}{J. Phys. B}
  \textbf{\bibinfo{volume}{35}}, \bibinfo{pages}{3095} (\bibinfo{year}{2002}).

\bibitem[{\citenamefont{Eiermann et~al.}(2003)\citenamefont{Eiermann,
  Treutlein, {Th. Anker}, Albiez, Taglieber, Marzlin, and
  Oberthaler}}]{EiermannTreutlein}
\bibinfo{author}{\bibfnamefont{B.}~\bibnamefont{Eiermann}},
  \bibinfo{author}{\bibfnamefont{P.}~\bibnamefont{Treutlein}},
  \bibinfo{author}{\bibnamefont{{Th. Anker}}},
  \bibinfo{author}{\bibfnamefont{M.}~\bibnamefont{Albiez}},
  \bibinfo{author}{\bibfnamefont{M.}~\bibnamefont{Taglieber}},
  \bibinfo{author}{\bibfnamefont{K.-P.} \bibnamefont{Marzlin}},
  \bibnamefont{and} \bibinfo{author}{\bibfnamefont{M.~K.}
  \bibnamefont{Oberthaler}}, \bibinfo{journal}{Phys. Rev. Lett.}
  \textbf{\bibinfo{volume}{91}}, \bibinfo{pages}{060402}
  (\bibinfo{year}{2003}).

\bibitem[{\citenamefont{Fallani et~al.}(2003)\citenamefont{Fallani, Cataliotti,
  Catani, Fort, Modugno, Zawada, and Inguscio}}]{FallaniCataliotti}
\bibinfo{author}{\bibfnamefont{L.}~\bibnamefont{Fallani}},
  \bibinfo{author}{\bibfnamefont{F.~S.} \bibnamefont{Cataliotti}},
  \bibinfo{author}{\bibfnamefont{J.}~\bibnamefont{Catani}},
  \bibinfo{author}{\bibfnamefont{C.}~\bibnamefont{Fort}},
  \bibinfo{author}{\bibfnamefont{M.}~\bibnamefont{Modugno}},
  \bibinfo{author}{\bibfnamefont{M.}~\bibnamefont{Zawada}}, \bibnamefont{and}
  \bibinfo{author}{\bibfnamefont{M.}~\bibnamefont{Inguscio}},
  \bibinfo{journal}{Phys. Rev. Lett.} \textbf{\bibinfo{volume}{91}},
  \bibinfo{pages}{240405} (\bibinfo{year}{2003}).

\bibitem[{\citenamefont{{Th. Anker} et~al.}(2004)\citenamefont{{Th. Anker},
  Albiez, Eiermann, Taglieber, and Oberthaler}}]{AnkerAlbiez}
\bibinfo{author}{\bibnamefont{{Th. Anker}}},
  \bibinfo{author}{\bibfnamefont{M.}~\bibnamefont{Albiez}},
  \bibinfo{author}{\bibfnamefont{B.}~\bibnamefont{Eiermann}},
  \bibinfo{author}{\bibfnamefont{M.}~\bibnamefont{Taglieber}},
  \bibnamefont{and} \bibinfo{author}{\bibfnamefont{M.~K.}
  \bibnamefont{Oberthaler}}, \bibinfo{journal}{Opt. Express}
  \textbf{\bibinfo{volume}{12}}, \bibinfo{pages}{11} (\bibinfo{year}{2004}).

\bibitem[{\citenamefont{Pu et~al.}(2003)\citenamefont{Pu, Baksmaty, Zhang,
  Bigelow, and Meystre}}]{Pu}
\bibinfo{author}{\bibfnamefont{H.}~\bibnamefont{Pu}},
  \bibinfo{author}{\bibfnamefont{L.~O.} \bibnamefont{Baksmaty}},
  \bibinfo{author}{\bibfnamefont{W.}~\bibnamefont{Zhang}},
  \bibinfo{author}{\bibfnamefont{N.~P.} \bibnamefont{Bigelow}},
  \bibnamefont{and} \bibinfo{author}{\bibfnamefont{P.}~\bibnamefont{Meystre}},
  \bibinfo{journal}{Phys. Rev. A} \textbf{\bibinfo{volume}{67}},
  \bibinfo{pages}{043605} (\bibinfo{year}{2003}).

\bibitem[{\citenamefont{Eiermann et~al.}(2004)\citenamefont{Eiermann, {Th.
  Anker}, Albiez, Taglieber, Treutlein, Marzlin, and
  Oberthaler}}]{EiermannAnker}
\bibinfo{author}{\bibfnamefont{B.}~\bibnamefont{Eiermann}},
  \bibinfo{author}{\bibnamefont{{Th. Anker}}},
  \bibinfo{author}{\bibfnamefont{M.}~\bibnamefont{Albiez}},
  \bibinfo{author}{\bibfnamefont{M.}~\bibnamefont{Taglieber}},
  \bibinfo{author}{\bibfnamefont{P.}~\bibnamefont{Treutlein}},
  \bibinfo{author}{\bibfnamefont{K.-P.} \bibnamefont{Marzlin}},
  \bibnamefont{and} \bibinfo{author}{\bibfnamefont{M.~K.}
  \bibnamefont{Oberthaler}}, \bibinfo{journal}{Phys. Rev. Lett.}
  \textbf{\bibinfo{volume}{92}}, \bibinfo{pages}{230401}
  (\bibinfo{year}{2004}).

\bibitem[{\citenamefont{Feng}(1993)}]{Feng}
\bibinfo{author}{\bibfnamefont{J.}~\bibnamefont{Feng}}, \bibinfo{journal}{Opt.
  Lett.} \textbf{\bibinfo{volume}{18}}, \bibinfo{pages}{1302}
  (\bibinfo{year}{1993}).

\bibitem[{\citenamefont{Yulin et~al.}(2002)\citenamefont{Yulin, Skryabin, and
  Firth}}]{Skryabin}
\bibinfo{author}{\bibfnamefont{A.~V.} \bibnamefont{Yulin}},
  \bibinfo{author}{\bibfnamefont{D.~V.} \bibnamefont{Skryabin}},
  \bibnamefont{and} \bibinfo{author}{\bibfnamefont{W.~J.} \bibnamefont{Firth}},
  \bibinfo{journal}{Phys. Rev. E} \textbf{\bibinfo{volume}{66}},
  \bibinfo{pages}{046603} (\bibinfo{year}{2002}).

\bibitem[{\citenamefont{Sukhorukov and {Yu S.
  Kivshar}}(2003)}]{SukhorukovKivshar}
\bibinfo{author}{\bibfnamefont{A.~A.} \bibnamefont{Sukhorukov}}
  \bibnamefont{and} \bibinfo{author}{\bibnamefont{{Yu S. Kivshar}}},
  \bibinfo{journal}{Opt. Lett.} \textbf{\bibinfo{volume}{28}},
  \bibinfo{pages}{2345} (\bibinfo{year}{2003}).

\bibitem[{\citenamefont{Mandelik et~al.}(2004)\citenamefont{Mandelik,
  Morandotti, Aitchison, and Silberberg}}]{Mandelik}
\bibinfo{author}{\bibfnamefont{D.}~\bibnamefont{Mandelik}},
  \bibinfo{author}{\bibfnamefont{R.}~\bibnamefont{Morandotti}},
  \bibinfo{author}{\bibfnamefont{J.~S.} \bibnamefont{Aitchison}},
  \bibnamefont{and}
  \bibinfo{author}{\bibfnamefont{Y.}~\bibnamefont{Silberberg}},
  \bibinfo{journal}{Phys. Rev. Lett.} \textbf{\bibinfo{volume}{92}},
  \bibinfo{pages}{093904} (\bibinfo{year}{2004}).

\bibitem[{\citenamefont{Neshev et~al.}(2004)\citenamefont{Neshev, Sukhorukov,
  Hanna, Krolikowski, and {Yu. S. Kivshar}}}]{Drago}
\bibinfo{author}{\bibfnamefont{D.}~\bibnamefont{Neshev}},
  \bibinfo{author}{\bibfnamefont{A.~A.} \bibnamefont{Sukhorukov}},
  \bibinfo{author}{\bibfnamefont{B.}~\bibnamefont{Hanna}},
  \bibinfo{author}{\bibfnamefont{W.}~\bibnamefont{Krolikowski}},
  \bibnamefont{and} \bibinfo{author}{\bibnamefont{{Yu. S. Kivshar}}},
  \bibinfo{journal}{Phys. Rev. Lett.} \textbf{\bibinfo{volume}{93}},
  \bibinfo{pages}{083905} (\bibinfo{year}{2004}).

\bibitem[{\citenamefont{Burger et~al.}(2001)\citenamefont{Burger, Cataliotti,
  Fort, Minardi, Inguscio, Chiofalo, and Tosi}}]{Burger}
\bibinfo{author}{\bibfnamefont{S.}~\bibnamefont{Burger}},
  \bibinfo{author}{\bibfnamefont{F.~S.} \bibnamefont{Cataliotti}},
  \bibinfo{author}{\bibfnamefont{C.}~\bibnamefont{Fort}},
  \bibinfo{author}{\bibfnamefont{F.}~\bibnamefont{Minardi}},
  \bibinfo{author}{\bibfnamefont{M.}~\bibnamefont{Inguscio}},
  \bibinfo{author}{\bibfnamefont{M.~L.} \bibnamefont{Chiofalo}},
  \bibnamefont{and} \bibinfo{author}{\bibfnamefont{M.~P.} \bibnamefont{Tosi}},
  \bibinfo{journal}{Phys. Rev. Lett.} \textbf{\bibinfo{volume}{86}},
  \bibinfo{pages}{4447} (\bibinfo{year}{2001}).

\bibitem[{\citenamefont{Peil et~al.}(2003)\citenamefont{Peil, Porto, {B.
  Laburthe Tolra}, Obrecht, King, Subbotin, Rolston, and Phillips}}]{PeilPorto}
\bibinfo{author}{\bibfnamefont{S.}~\bibnamefont{Peil}},
  \bibinfo{author}{\bibfnamefont{J.~V.} \bibnamefont{Porto}},
  \bibinfo{author}{\bibfnamefont{B.~L.} \bibnamefont{Tolra}},
  \bibinfo{author}{\bibfnamefont{J.~M.} \bibnamefont{Obrecht}},
  \bibinfo{author}{\bibfnamefont{B.~E.} \bibnamefont{King}},
  \bibinfo{author}{\bibfnamefont{M.}~\bibnamefont{Subbotin}},
  \bibinfo{author}{\bibfnamefont{S.~L.} \bibnamefont{Rolston}},
  \bibnamefont{and} \bibinfo{author}{\bibfnamefont{W.~D.}
  \bibnamefont{Phillips}}, \bibinfo{journal}{Phys. Rev. A}
  \textbf{\bibinfo{volume}{67}}, \bibinfo{pages}{051603}
  (\bibinfo{year}{2003}).

\bibitem[{\citenamefont{Perez-Garcia et~al.}(1998)\citenamefont{Perez-Garcia, Michinel,
  and Herrero}}]{Perez-Garcia}
\bibinfo{author}{\bibfnamefont{V.~M.} \bibnamefont{P\'{e}rez-Garc\'{i}a}},
  \bibinfo{author}{\bibfnamefont{H.} \bibnamefont{Michinel}},
  \bibnamefont{and} \bibinfo{author}{\bibfnamefont{H.}
  \bibnamefont{Herrero}}, \bibinfo{journal}{Phys. Rev. A}
  \textbf{\bibinfo{volume}{57}}, \bibinfo{pages}{3837} (\bibinfo{year}{1998}).

\bibitem[{\citenamefont{Grynberg and Robilliard}(2001)}]{GrynbergRobilliard}
\bibinfo{author}{\bibfnamefont{G.}~\bibnamefont{Grynberg}} \bibnamefont{and}
  \bibinfo{author}{\bibfnamefont{C.}~\bibnamefont{Robilliard}},
  \bibinfo{journal}{Phys. Rep.} \textbf{\bibinfo{volume}{355}},
  \bibinfo{pages}{335} (\bibinfo{year}{2001}).

\bibitem[{\citenamefont{Fort et~al.}(2003)\citenamefont{Fort, Cataliotti,
  Fallani, Ferlaino, Maddaloni, and Inguscio}}]{FortCataliotti}
\bibinfo{author}{\bibfnamefont{C.}~\bibnamefont{Fort}},
  \bibinfo{author}{\bibfnamefont{F.~S.} \bibnamefont{Cataliotti}},
  \bibinfo{author}{\bibfnamefont{L.}~\bibnamefont{Fallani}},
  \bibinfo{author}{\bibfnamefont{F.}~\bibnamefont{Ferlaino}},
  \bibinfo{author}{\bibfnamefont{P.}~\bibnamefont{Maddaloni}},
  \bibnamefont{and} \bibinfo{author}{\bibfnamefont{M.}~\bibnamefont{Inguscio}},
  \bibinfo{journal}{Phys. Rev. Lett.} \textbf{\bibinfo{volume}{90}},
  \bibinfo{pages}{140405} (\bibinfo{year}{2003}).

\bibitem[{\citenamefont{de~Sterke and Sipe}(1988)}]{Sipe}
\bibinfo{author}{\bibfnamefont{C.~M.} \bibnamefont{de~Sterke}}
  \bibnamefont{and} \bibinfo{author}{\bibfnamefont{J.~E.} \bibnamefont{Sipe}},
  \bibinfo{journal}{Phys. Rev. A} \textbf{\bibinfo{volume}{38}},
  \bibinfo{pages}{5149} (\bibinfo{year}{1988}).

\bibitem[{\citenamefont{Steel and Zhang}(1998)}]{SteelZhang}
\bibinfo{author}{\bibfnamefont{M.~J.} \bibnamefont{Steel}} \bibnamefont{and}
  \bibinfo{author}{\bibfnamefont{W.~P.} \bibnamefont{Zhang}}
  (\bibinfo{year}{1998}), \bibinfo{note}{cond-mat/9810284}.

\bibitem[{\citenamefont{Zobay et~al.}(1999)\citenamefont{Zobay, P\"otting,
  Meystre, and Wright}}]{Zobay}
\bibinfo{author}{\bibfnamefont{O.}~\bibnamefont{Zobay}},
  \bibinfo{author}{\bibfnamefont{S.}~\bibnamefont{P\"otting}},
  \bibinfo{author}{\bibfnamefont{P.}~\bibnamefont{Meystre}}, \bibnamefont{and}
  \bibinfo{author}{\bibfnamefont{E.~M.} \bibnamefont{Wright}},
  \bibinfo{journal}{Phys. Rev. A} \textbf{\bibinfo{volume}{59}},
  \bibinfo{pages}{643} (\bibinfo{year}{1999}).

\bibitem[{\citenamefont{Alfimov et~al.}(2002)}]{Alfimov}
\bibinfo{author}{\bibfnamefont{G.~L.} \bibnamefont{Alfimov}},
\bibinfo{author}{\bibfnamefont{V.~V.} \bibnamefont{Konotop}},
  \bibnamefont{and} \bibinfo{author}{\bibfnamefont{M.}
  \bibnamefont{Salerno}}, \bibinfo{journal}{Europhys. Lett.}
  \textbf{\bibinfo{volume}{58}}, \bibinfo{pages}{7}
  (\bibinfo{year}{2002}).

\bibitem[{\citenamefont{Louis et~al.}(2003)\citenamefont{Louis, Ostrovskaya,
  Savage, and {Yu. S. Kivshar}}}]{Pearl}
\bibinfo{author}{\bibfnamefont{P.~J.~Y.} \bibnamefont{Louis}},
  \bibinfo{author}{\bibfnamefont{E.~A.} \bibnamefont{Ostrovskaya}},
  \bibinfo{author}{\bibfnamefont{C.~M.} \bibnamefont{Savage}},
  \bibnamefont{and} \bibinfo{author}{\bibnamefont{{Yu. S. Kivshar}}},
  \bibinfo{journal}{Phys. Rev. A} \textbf{\bibinfo{volume}{67}},
  \bibinfo{pages}{013602} (\bibinfo{year}{2003}).

\bibitem[{\citenamefont{Efremidis and Christodoulides}(2003)}]{Efremidis}
\bibinfo{author}{\bibfnamefont{N.~K.} \bibnamefont{Efremidis}}
  \bibnamefont{and} \bibinfo{author}{\bibfnamefont{D.~N.}
  \bibnamefont{Christodoulides}}, \bibinfo{journal}{Phys. Rev. A}
  \textbf{\bibinfo{volume}{67}}, \bibinfo{pages}{063608}
  (\bibinfo{year}{2003}).

\bibitem[{\citenamefont{Pelinovsky et~al.}()\citenamefont{Pelinovsky,
  Sukhorukov, and {Yu. S. Kivshar}}}]{Dima}
\bibinfo{author}{\bibfnamefont{D.~E.} \bibnamefont{Pelinovsky}},
  \bibinfo{author}{\bibfnamefont{A.~A.} \bibnamefont{Sukhorukov}},
  \bibnamefont{and} \bibinfo{author}{\bibnamefont{{Yu. S. Kivshar}}},
 \bibinfo{journal}{Phys. Rev. E} \textbf{\bibinfo{volume}{70}},
  \bibinfo{pages}{036618} (\bibinfo{year}{2004}).

\bibitem[{\citenamefont{Morsch et~al.}(2002)\citenamefont{Morsch, Cristiani, Muller,
  Ciampini, and Arimondo}}]{morsch}
\bibinfo{author}{\bibfnamefont{O.}~\bibnamefont{Morsch}},
  \bibinfo{author}{\bibfnamefont{M.} \bibnamefont{Cristiani}},
  \bibinfo{author}{\bibfnamefont{J.~H.}~\bibnamefont{M\"uller}},
  \bibinfo{author}{\bibfnamefont{D.}~\bibnamefont{Ciampini}}, \bibnamefont{and}
  \bibinfo{author}{\bibfnamefont{E.}~\bibnamefont{Arimondo}},
  \bibinfo{journal}{Phys. Rev. A} \textbf{\bibinfo{volume}{66}},
  \bibinfo{pages}{021601(R)} (\bibinfo{year}{2002}).

\bibitem[{\citenamefont{Adhikari}(2003)\citenamefont{Adhikari}}]{adhikari}
\bibinfo{author}{\bibfnamefont{S. K.}~\bibnamefont{Adhikari}},
  \bibinfo{journal}{J. Phys. B} \textbf{\bibinfo{volume}{36}},
  \bibinfo{pages}{3951} (\bibinfo{year}{2003}).

\bibitem[{\citenamefont{Fromhold et~al.}(2000)\citenamefont{Fromhold, Tench, Bujkiewicz, Wilkinson, Sheard}}]{chaos}
\bibinfo{author}{\bibfnamefont{T.~M.}~\bibnamefont{Fromhold}},
  \bibinfo{author}{\bibfnamefont{C.~R.} \bibnamefont{Tench}},
  \bibinfo{author}{\bibfnamefont{S.}~\bibnamefont{Bujkiewicz}},
  \bibinfo{author}{\bibfnamefont{P.~B.}~\bibnamefont{Wilkinson}}, \bibnamefont{and}
  \bibinfo{author}{\bibfnamefont{F.~W.}~\bibnamefont{Sheard}},
  \bibinfo{journal}{J. Opt. B} \textbf{\bibinfo{volume}{2}},
  \bibinfo{pages}{628} (\bibinfo{year}{2000}).

\bibitem[{\citenamefont{Peik et~al.}(1997)\citenamefont{Peik, Dahan, Bouchoule,
  Castin, and Salomon}}]{PeikBenDahn}
\bibinfo{author}{\bibfnamefont{E.}~\bibnamefont{Peik}},
  \bibinfo{author}{\bibfnamefont{M.} \bibnamefont{Ben Dahan}},
  \bibinfo{author}{\bibfnamefont{I.}~\bibnamefont{Bouchoule}},
  \bibinfo{author}{\bibfnamefont{Y.}~\bibnamefont{Castin}}, \bibnamefont{and}
  \bibinfo{author}{\bibfnamefont{C.}~\bibnamefont{Salomon}},
  \bibinfo{journal}{Phys. Rev. A} \textbf{\bibinfo{volume}{55}},
  \bibinfo{pages}{2989} (\bibinfo{year}{1997}).

\bibitem[{\citenamefont{Dudarev et~al.}(2004)\citenamefont{Dudarev, Diener, and
  Niu}}]{DudarevDiener}
\bibinfo{author}{\bibfnamefont{A.~M.} \bibnamefont{Dudarev}},
  \bibinfo{author}{\bibfnamefont{R.~B.} \bibnamefont{Diener}},
  \bibnamefont{and} \bibinfo{author}{\bibfnamefont{Q.}~\bibnamefont{Niu}},
  \bibinfo{journal}{J. Opt. B} \textbf{\bibinfo{volume}{6}},
  \bibinfo{pages}{S231} (\bibinfo{year}{2004}).

\bibitem[{\citenamefont{Agrawal}(2001)}]{Agrawal}
\bibinfo{author}{\bibfnamefont{G.~P.} \bibnamefont{Agrawal}},
  \emph{\bibinfo{title}{Nonlinear Fiber Optics}} (\bibinfo{publisher}{Academic
  Press}, \bibinfo{address}{San Diego}, \bibinfo{year}{2001}),
  \bibinfo{edition}{3rd} ed.

\bibitem[{\citenamefont{D.Burak and Nasalski}(1994)}]{Burak}
\bibinfo{author}{\bibnamefont{D.Burak}} \bibnamefont{and}
  \bibinfo{author}{\bibfnamefont{W.}~\bibnamefont{Nasalski}},
  \bibinfo{journal}{Appl. Opt} \textbf{\bibinfo{volume}{33}},
  \bibinfo{pages}{6393} (\bibinfo{year}{1994}).

\bibitem[{xmd()}]{xmds}
\bibinfo{note}{The xmds website is at http://www.xmds.org}.

\end{thebibliography}
\end{document}